\documentstyle[preprint,aps,prl]{revtex} 
\begin{document}
\title{Coulomb interactions at quantum Hall critical points\\
of systems in a periodic potential}
\author{Jinwu Ye}
\address{Department of Physics,
The Johns Hopkins University, Baltimore MD, 21218}
\author{Subir Sachdev}
\address{Department of Physics, P.O. Box 208120,
Yale University, New Haven, CT 06520-8120}
\date{\today}
\maketitle
\begin{abstract}
We study the consequences of long-range Coulomb interactions at
the critical points between integer/fractional quantum Hall states
and an insulator. We use low energy theories for such transitions
in anyon gases in the presence of an external periodic potential.
We find that Coulomb interactions are marginally irrelevant for
the integer quantum Hall case. For the fractional case, depending upon
the anyon statistics parameter, we find
behavior similar to the integer case, or flow to a novel line of fixed
points with exponents $z=1$, $\nu > 1$ stable against weak disorder in the 
position of the critical point,
or run-away flow to strong coupling.
\end{abstract}
\narrowtext

The zero temperature quantum phase transitions between the different quantum Hall
and insulating states of a two-dimensional electron gas in a strong magnetic field 
are among the most intensively studied quantum critical points, 
both theoretically~\cite{bodo,shahar} and experimentally~\cite{expts}.
Earlier theoretical investigations focussed on the transitions
between the integer quantum Hall plateaus and 
described them in terms of non-interacting electrons moving 
in a random external potential~\cite{ammp}. It has also been
argued that the transitions between fractional quantum Hall
states could be mapped onto models essentially equivalent
to those between the integer states~\cite{jkt}. The latter point of
view was however questioned by Wen and Wu~\cite{wen} and
Chen, Fisher and Wu~\cite{chen}:
they focussed on the simpler case of systems in the presence of
a {\em periodic} rather than a random potential, and examined a model
of anyons, with a statistical angle $\theta$ and short-range repulsive
interactions, which displayed a second order quantum phase transition
between a quantized Hall state and a Mott insulator as the strength of the
periodic potential was varied.
This transition was characterized by a line of critical points
with continuously varying exponents, parametrized by the value
of $\theta$.
For the case $\theta = 0$, when the anyons were fermions, 
the transition was out of a integer quantum Hall state; its exponents and
other universal properties were different from the cases
$0 < \theta < 2 \pi$ for which the anyons acquired fractional statistics
and the transition was out a fractional quantum Hall state. 
(For $\theta = 2 \pi$ the anyons became bosons and the Hall
state reduced to a superfluid.)

In all of the above theoretical works, the long-range 
Coulomb interactions between charge carriers have been effectively ignored.
However, a few recent works have taken steps to remedy this
serious shortcoming. Yang {\em et al.}~\cite{yang} studied the integer
quantum Hall transition under a Hatree-Fock treatment of the Coulomb
interaction. Lee and Wang~\cite{lee} showed that the renormalization
group eigenvalue of the Coulomb interaction was zero at the Hartree-Fock 
critical point; higher order calculations are therefore
necessary to understand the physics.
Pfannkuche and MacDonald~\cite{allan} numerically studied
electrons with Coulomb interactions in a periodic potential 
between a fractional Hall state
and an insulator, but were limited to rather small system sizes.
Interesting scaling interpretations of Coulomb interaction-induced
dephasing were discussed in Ref~\cite{polya}.

In this paper we shall provide a thorough analysis of the consquences
of Coulomb interactions on the anyons in a periodic potential model
of Refs~\cite{wen,chen}. We shall show that the Coulomb interaction
is {\em marginally irrelevant} for the integer case ($\theta=0$), and remains
so for the fractional case for small values of $\theta$; this marginally irrelevant
interaction will lead to logarithmic corrections to naive scaling functions for
the vicinity of the transition.
For larger $\theta$, we will establish, in a certain $1/N$ expansion, the
existence of a novel line of fixed points at which the Coulomb interactions
acquire a non-zero fixed point value determined by the value of $\theta$. 
There are no logarithmic corrections at these fixed points, and naive scaling holds.
We find a dynamic critical exponent $z=1$ at all points on the fixed line, providing
a concrete realization of the scenario~\cite{mpaf,shahar}, 
not previously established explicitly,
that energies must scale as inverse distances for the $1/r$ Coulomb interaction.
It is also worth noting that, despite the value $z=1$, the critical
correlators are not Lorentz invariant.
We also find that the correlation length exponent $\nu$ satisfies
$\nu > 2/d$ (where $d=2$ is the spatial dimensionality) along this fixed line,
which implies that the fixed line is stable towards disorder involving local
fluctuations in the position of the critical point.

We begin our analysis by writing down the model of Ref~\cite{chen}
extended to include Coulomb interactions between the charge carriers
\begin{eqnarray}
&& {\cal L}  = \int d^d x d \tau \Biggl[ \alpha  \overline{\psi}_{m} \gamma_{0} \partial_{0}
    \psi_{m} + \overline{\psi}_{m} \gamma_{i} \partial_{i} \psi_{m} \nonumber\\
&&~~~~~~~~~~~~~-\frac{i}{\sqrt{N}} q \mu^{\epsilon/2} \alpha^{1/2} a_{0}
\overline{\psi}_{m} \gamma_{0} \psi_{m}
-\frac{i}{\sqrt{N}} g \mu^{\epsilon/2} \alpha^{1/2} a_{i}
\overline{\psi}_{m} \gamma_{i} \psi_{m} \Biggr] \nonumber\\
&&~~~~~~~~~~~~ + \int \frac{d^2 k}{4 \pi^2} \frac{d \omega}{2 \pi}
\left[i k a_{0}(-\vec{k},-\omega) a_{t}(\vec{k}, \omega)
+ \frac{k}{2} a_{t}(-\vec{k},-\omega) a_{t}(\vec{k}, \omega) \right]
\label{classical}
\end{eqnarray}
The $\psi_m$ are $m=1\cdots N$ species of charge $q/\sqrt{N}$
2+1 dimensional Dirac fermions which interact with a $U(1)$ gauge field
$(a_0, a_i)$ ($i=1,2$); we are interested in the case $N=1$ but will find the
large $N$ expansion to be a useful tool. The $\gamma_0, \gamma_i$ are the Dirac
$\gamma$ matrices, $x_i$ ($\tau$) are spatial (temporal) co-ordinates
with $\partial_0 \equiv \partial_{\tau}$, $\partial_i \equiv \partial_{x_i}$,
and $\vec{k}$, $\omega$ ($k = |\vec{k}|$)
are the Fourier transformed wavevector and frequency
variables.
To aid the subsequent renormalization group analysis, we are working in $d=2+\epsilon$
spatial dimensions and $\mu$ is a renormalization scale. 
The parameter $\alpha$ is introduced to allow for anisotropic renormalization
between space and time~\cite{cardy}.
We have used the Coulomb
gauge which allows us to explicitly represent $a_i$ in terms of the
transverse spatial component with $a_i = i \epsilon_{ij} k_j a_t / k$.
The term before last in ${\cal L}$ is the Chern Simons coupling:
it turns the Dirac  particles into anyons with a statistical angle
$\theta/N$ with $\theta \equiv qg$; notice
 that the angle is of order $1/N$ and so the expected periodicity of the
physics under $\theta/N \rightarrow \theta/N +4\pi$ will not be visible in the $1/N$ expansion. 
The last term is the Coulomb interaction,
and it has been written in terms of $a_t$ following Ref~\cite{bert}.

In the absence of the last Coulomb interaction term, it was shown in 
Ref~\cite{chen} that ${\cal L}$ represents the critical theory of a system
of anyons in a periodic potential undergoing a transition from an insulator
with conductivities $\sigma_{xx} = \sigma_{xy} = 0$ into a fractional
quantum Hall state with $\sigma_{xx} = 0$ and $\sigma_{xy} = (q^2 /h)/(1- \theta/2 \pi)$
to leading order in $1/N$.
Both these states have energy gaps, and we have shown that the Coulomb
interaction does not modify the values of $\sigma_{ij}$ in either phase.
The relationship of the continuum model ${\cal L}$ to the more realistic model
of electrons studied in Ref~\cite{allan} remains somewhat unclear, although it is 
plausible that ${\cal L}$ is the critical theory of the latter. We may also 
view ${\cal L}$ as the simplest theory consistent with the following requirements,
and therefore worthy of further study: ({\em i\/}) the two phases on either side of
the critical point have the correct values of $\sigma_{ij}$, and the Hall phase
has {\em both\/} quasi-particle and quasi-hole excitations with the correct
charge and statistics, and
({\em ii\/}) the gap towards the quasi-particle {\em and\/} the quasi-hole
excitations vanishes at the critical point.

We now proceed with a renormalization group analysis of ${\cal L}$. 
Simple power counting shows that both the Chern-Simons and Coulomb interactions~\cite{subir}
are marginal at tree level in $d=2$, and so loop expansions are required 
and useful. Power counting also shows that a short-range four-fermion interaction
term is {\em irrelevant} and has therefore been neglected in ${\cal L}$; this makes
the fermionic formulation of the anyon problem much simpler than its
bosonic counterpart~\cite{wen,fisher,long}.

The loop expansion requires counterterms to account for ultraviolet divergences
in momentum integrals; we write the counter terms as
\begin{eqnarray}
&& {\cal L}  = \int d^2 x d \tau \Biggl[ \alpha (Z_{\alpha} - 1)
 \overline{\psi}_{m} \gamma_{0} \partial_{0}
    \psi_{m} + (Z_2 -1 )\overline{\psi}_{m} \gamma_{i} \partial_{i} \psi_{m} \nonumber\\
&&~~~~~~~~~~~~~-\frac{i}{\sqrt{N}} (Z_1^{q} - 1) 
q \mu^{\epsilon/2} \alpha^{1/2} a_{0}
\overline{\psi}_{m} \gamma_{0} \psi_{m}
-\frac{i}{\sqrt{N}} (Z_1^{g} - 1) g \mu^{\epsilon/2} \alpha^{1/2} a_{i}
\overline{\psi}_{m} \gamma_{i} \psi_{m} \Biggr] 
\label{ct}
\end{eqnarray}
In general, counter terms for the last two gauge field terms in ${\cal L}$ should also be 
considered.
However, we have shown~\cite{long} that at least to two loops, there no divergences
associated with these terms. The Ward identities following from gauge invariance
dictate $Z_1^q = Z_{\alpha}$ and $Z_{1}^{g} = Z_2$. Using these identities
we relate the bare fields and couplings in ${\cal L}$ to the renormalized
quantities by 
$\psi_{mB} = Z_2^{1/2} \psi_m$, $\alpha_B = (Z_{\alpha}/Z_2 ) \alpha$, 
$q_B = q \mu^{\epsilon} (Z_{\alpha} / Z_2 )^{1/2}$ and $g_B = 
g \mu^{\epsilon/2} (Z_2 / Z_{\alpha} )^{1/2}$.
Notice that these relations imply that for the statistical angle $\theta/N = qg/N$ we have
$\theta_B = \theta \mu^{\epsilon}$; so in $d=2$ this angle is a renormalization group
invariant, which is expected on general physical grounds. 
The dynamic critical exponent, $z$ is related to the renormalization of $\alpha$
by~\cite{cardy}
\begin{equation}
z = 1 - \mu \frac{d}{d\mu} \ln \alpha = 1 - \mu \frac{d}{d\mu} \ln \frac{Z_2}{Z_{\alpha}}
\label{zval}
\end{equation}
We will find it convenient to express the loop expansion in terms of 
the ``fine structure'' constant $w \equiv \pi q^2 / 8$, and a central object of
study shall be its $\beta$-function $\beta ( w ) = \mu (dw/d\mu)$. By comparing
(\ref{zval}) with relationships between bare and renormalized quantities quoted above
we see that
\begin{equation}
z = 1 - \beta (w)/w.
\label{zbeta}
\end{equation}
Finally, the critical exponent $\nu$ is related to the anomalous dimension
of the composite operator $\overline{\psi} \psi$ by $\nu^{-1} - 1
= \mu (d \ln Z_{\overline{\psi} \psi} / d \mu)$; the renormalization constant
$Z_{\overline{\psi} \psi}$ can be calculated by inserting the 
operator into the self-energy diagrams.

We begin the explicit calculation of the renormalization constants by considering
a direct perturbative expansion in the Coulomb fine structure constant $w$
and the statistical andgle $\theta$. At one-loop order, we find no dependence
on $\theta$; the values of the renormalization constants upto terms of order
$w^2$, $\theta^2$ and $w \theta$ are
\begin{equation}
Z_2 = 1 - 2 w /N \pi \epsilon , ~~~~Z_{\alpha} = 1,~~~~~Z_{\overline{\psi} \psi} = Z_2.
\label{oneloopz}
\end{equation}
We also explicitly verified that the gauge invariance Ward identities hold. From these
results we find for the $\beta$-function of the Coulomb coupling
\begin{equation}
\beta (w ) = \frac{2 w^2}{N \pi} + {\cal O}( w^3 , w^2 \theta^2 )
\label{oneloopbeta}
\end{equation}
while the critical exponents are
\begin{equation}
z = 1 - 2 w /N \pi,~~~~~\nu = 1 - 2 w /N \pi
\label{oneloopexp}
\end{equation}
to be evaluated at the fixed point of the $\beta$-function.
The result (\ref{oneloopbeta}) shows that the $w$ is marginally irrelevant
and flows to the fixed point $w^{\ast} = 0$ at long distances. During this flow
(\ref{oneloopexp}) shows that the effective $z < 1$, corresponding to a smaller
density of states at low energies, which is physically consistent with
the irrelevance of Coulomb interactions. For the integer Hall state case
we have $\theta =0$, and then the fixed point is simply a free Dirac theory:
in this case the Coulomb interactions are {\em dangerously\/} irrelevant, as it is responsible
for the $T$ dependence of physical quantities and will lead to logarithmic corrections
to naive scaling.

To understand larger values of $\theta$, and to explore the consequences of a 
possible interference between the Coulomb interactions and the Chern-Simons term
we found it convenient to perform a $1/N$ expansion. This is technically
simpler than a perturbative two-loop extension of the computation above,
and also automatically includes the dynamic screening of the gauge field propagator
by the fermion polarization~\cite{bert}. Alternatively stated, the so-called RPA
approximation becomes exact at $N=\infty$, and $1/N$ corrections require
gauge field propagators which have the RPA form
\begin{equation}
{\cal L}_{RPA} = \frac{1}{2} \int \frac{d^2 k}{4 \pi^2} \frac{d \omega}{2 \pi}
(a_0 , a_t)
\left( \begin{array}{cc}
q^2 k^2 /(16 \sqrt{k^2 + \omega^2}) & i k \\
i k & k + g^2 /(16 \sqrt{k^2 + \omega^2}) \end{array}
\right) \left( \begin{array}{c} a_0 \\ a_t \end{array} \right)
\label{lrpa}
\end{equation}
Evaulating the fermion self-energy diagrams to order $ 1/N $, we find
for the renormalization constants
\begin{eqnarray}
 Z_{2} & = &1-\frac{1}{ N \epsilon} \left( \frac{2 w}{ \pi \lambda}
    -\frac{ 16 w^2 A}{\pi^{2} \lambda} + \frac{ \theta^{2} C}{ 16 \pi^{2}}
    -\frac{\theta^{2} E}{16 \pi^{2}} \right)    \nonumber   \\
 Z_{\alpha} & = &1-\frac{1}{ N \epsilon} \left(
    \frac{16 w^2 B}{\pi^{2} \lambda} - \frac{ \theta^{2} D}{ 16 \pi^{2}}
    +\frac{\theta^{2} F}{16 \pi^{2}} \right),
\label{diverge}
\end{eqnarray}
where $ \lambda=1+ (\theta/16)^{2}$ and the constants $ A,B,C,D, E=A+B, F=B $
are given by the formal expressions
\begin{eqnarray}
  A & = & \int^{1}_{0} d x \frac{ 4 x^{2} (1-x^{2}) }{ (1+x^{2})^{3} }
  f(x;w,\theta),  
  ~~~~~ B= \int^{1}_{0} d x \frac{ (1-x^{2})(1-6 x^{2}+x^{4}) }{ (1+x^{2})^{3} }
  f(x;w,\theta)     \nonumber \\
  C & = & \int^{1}_{0} d x \frac{ 4 x^{2} }{(1-x^{2})
  (1+x^{2}) } f(x;w,\theta),  
  ~~~~~ D= \int^{1}_{0} d x \frac{ (1-6 x^{2}+x^{4}) }
  { (1-x^{2}) (1+x^{2}) } f(x;w,\theta) 
\label{constant}
\end{eqnarray}
with $ f(x;w,\theta)=( \lambda (1+x^{2})+ w(1-x^{2}))^{-1} $, and the variable $x$
represents an intermediate frequency.
Note the two constants $ C, D $ are divergent: this 
divergence is due to the singular effect of frequencies $|\omega| \gg k$. 
However, as shown below and in Ref~\cite{long}, these divergences are gauge artifacts
and cancel in the $\beta$-function and in
   any physical gauge-invariant quantity like $ \nu$, $z $
  or $ \sigma_{ij}$. The divergences however do infect
the anomalous dimension of the field operator $\psi$:
this is as expected as the propagator of $\psi$ is clearly gauge-dependent.

We computed the $\beta$-function of the Coulomb coupling $w$ from the renormalization
constants (\ref{diverge}); the divergences do indeed cancel and we obtain the
result
\begin{eqnarray}
&& \beta (w) = \frac{2 w^{2} (1-\phi)}{ N \pi^{2} 
   \lambda^{2}} \left[ \pi-16 w \int^{1}_{0} d x \left(\frac{ 1-x^{2}}{1+x^{2}}
\right)^{3}
   \frac{ \lambda (1+x^{2})+ \frac{ w}{2}(1-x^{2}) }{( \lambda (1+x^{2})
   +w (1-x^{2}))^{2}} \right]  \nonumber   \\
    &&~~~~~~ +  \frac{32 w \phi}{ N \pi^{2}}
     \int^{1}_{0} d x \frac{( 1-x^{2})(-1+10 x^{2}-x^{4} )}{(1+x^{2})^{3}}
   \frac{ (1+x^{2})+ \frac{ w}{2}(1-x^{2}) }{( \lambda (1+x^{2})
   +w (1-x^{2}))^{2}},
\label{complete}
\end{eqnarray}
where $ \phi \equiv (\theta/16)^{2}$ and $\lambda=1+ \phi $. 
In a similar manner, the effective exponent $z$ is given by (\ref{zbeta}),
and for the exponent $\nu$ we obtain 
\begin{eqnarray}
&& \frac{1}{\nu} = 1 -\frac{128 \phi}{ N \pi^{2} } \int^{1}_{0} d x \frac{( 1-x^{2})
   (1+6 x^2+ x^{4})}{(1+x^{2})^{3}}
   \frac{ 1+x^{2}+ \frac{ w}{2}(1-x^{2}) }{( \lambda (1+x^{2})
   +w (1-x^{2}))^{2}}  \nonumber   \\
&&~~~~~~~~~~~ + \frac{512 \phi (1- \phi)}{ N \pi^{2}}
     \int^{1}_{0} d x \frac{( 1-x^{2})(1+ x^{2})}
   {( \lambda (1+x^{2}) +w (1-x^{2}))^{3}};
\label{disorder}
\end{eqnarray}
in this last expression we have used the fact that $\beta ( w ) =0$ at
a fixed point to simplify the result a bit.
In the absence of Coulomb interactions ($w=0$) the above result for $\nu$
becomes:
\begin{equation}
\nu=1-\frac{ 512 \phi (1-2 \phi) }{ N 3 \pi^{2} \lambda^{3} }
\end{equation}

  It agrees with earlier results~\cite{chen,wrong} obtained in a very different
computation in the Lorentz gauge: this agreement is another non-trivial check
of our renormalization procedure.

We now turn to the physical implications of our main
results (\ref{complete}) and (\ref{disorder}).

First, consider the transition out of the integer quantum Hall state, $\theta = 0$,
which implies $\phi=0$, $\lambda = 1$. Then simple analysis of (\ref{complete})
shows that $\beta (w) > 0$ for all $w > 0$; for small $w$ we have $\beta (w) =
2 w^2 / N \pi$, in agreement with one-loop result (\ref{oneloopbeta}), while for $w \gg 1$,
$\beta (w) = 4 / N \pi$. So the only fixed point remains at $w = 0$,
and the prediction of the large $N$ theory agrees with the weak-coupling 
analysis--Coulomb interactions are dangerously irrelevant. This agreement between the two
approaches is reassuring as it is not {\em a priori} required: it is absent in 
the bosonic formulation~\cite{wen} of the same critical point.

Turning to the fractional case with a non-zero $\theta$, we show a plot of 
a numerical integration of
the flows implied by (\ref{complete}) in Fig~1; there are three distinct
regimes:
\newline
({\em i}) $ \phi < \phi_{c1}$
\newline
For small values of $\phi$ the $w=0$ fixed point remains stable, as for the integer
case above. The limiting value $\phi_{c1}$ can be determined by expanding
$\beta(w)$ in (\ref{complete}) in powers of $w$:
\begin{equation}
\beta(w)= \frac{ w^{2}}{ 2 N \pi \lambda^{3}}(4+3\phi-5 \phi^{2})
      -\frac{ 32 w^{3}}{15 N  \pi^{3} \lambda^{4}}( 5+4 \phi-7 \phi^{2})
\end{equation}
The co-efficient of $w^2$ changes sign at
$ \phi= \phi_{c1} = (3+\sqrt{89})/10 \approx 1.24$, beyond which the $w=0$
point is no longer stable.
\newline
({\em ii}) $\phi_{c1} < \phi < \phi_{c2}$
\newline
For intermediate values of $\phi$, the flow is towards an attractive line
of fixed points $0< w^{\ast} (\phi) < \infty$. The value $\phi=\phi_{c2}$
at which $w^{\ast} ( \phi ) \rightarrow \infty$ can be determined by
evaluating (\ref{complete}) in the large $w$ limit:
\begin{equation}
\beta(w \rightarrow \infty )= \frac{ 2(2-\phi)}{  N \pi}.
\end{equation}
This shows that the flow is away from $w=\infty$ for $\phi < \phi_{c2} = 2$.
This line of stable fixed points for the present range of $\phi$ 
is our main new result. 
We can easily determine the values of the fixed-point critical exponents:
from (\ref{zbeta}) we see that the $z=1$, while $\nu$ follows from (\ref{disorder}).
We find $\nu (\phi_{c1} ) \approx 1 + 2.82/N$, and $\nu (\phi_{c2}^{-})=
1 + 1/N w^{\ast}$, and a monotonic change in between. 
As noted earlier, because $\nu > 1$, this line is stable to
disorder in the local position of the critical point.
\newline
({\em iii}) $\phi > \phi_{c2}$
\newline
Now the flow is to $w^{\ast} = \infty$. However, the flows cannot be trusted
once $w \sim N$, and so we are unable to draw any firm conclusions about this regime.
     
To conclude, we have presented an analysis of the consequences
of Coulomb interactions at quantum Hall critical points which goes well
beyond the linear stability/Hartree-Fock treatments in earlier
works~\cite{yang,lee}. To allow such a study,
we simplified the usual physical situation by replacing the random external
potential by a periodic one. Nevertheless, it is quite interesting that
we found a fixed line which is stable towards the introduction of small disorder
in the position of the critical point.

We thank M.P.A.~Fisher, B.~Halperin, M.~Franz, C.~Kane, S.~Kivelson, A.~Millis,
N.~Read, R.~Shankar, Z.~Wang and Y. S. Wu for helpful discussions.
This work was initiated 
at the Aspen Center for Physics and supported by
NSF Grant No. DMR-97-07701 (J.Ye) and DMR-96-23181 (S.S.)

\end{document}